\documentclass[preprint]{aastex}
\usepackage{emulateapj5}
\usepackage{apjfonts}

\shorttitle{Unidentified Gamma-ray Sources}
\shortauthors{Kawasaki \& Totani}

\begin{document}

%----------------------------------------------------------------------

\title{Positional Coincidence between the High-latitude Steady Unidentified
Gamma-ray Sources and Possibly Merging Clusters of Galaxies}

\author{Wataru Kawasaki\altaffilmark{1,2,3} and 
Tomonori Totani\altaffilmark{4,5}}

%\email{kawasaki@astron.s.u-tokyo.ac.jp, totani@th.nao.ac.jp}
\slugcomment{ApJ, Accepted 2002 Apr 19; Submitted 2001 Aug 19}

\altaffiltext{1}{Department of Astronomy, The University of Tokyo, 
7-3-1, Hongo, Bunkyo-ku, Tokyo, 113-0033, Japan; 
kawasaki@astron.s.u-tokyo.ac.jp}

\altaffiltext{2}{JSPS Postdoctoral Fellow.}

\altaffiltext{3}{Present Address: 
Institute of Astronomy and Astrophysics, Academia Sinica, 
P.O.Box 23-141, Taipei, 106, Taiwan, Republic of China}

\altaffiltext{4}{Princeton University Observatory, 
Peyton Hall, Princeton, 08544-1001, NJ, USA}

\altaffiltext{5}{Theory Division, National Astronomical Observatory, 
Mitaka, Tokyo, 181-8588, Japan; 
totani@th.nao.ac.jp}

%----------------------------------------------------------------------

\begin{abstract}
We report an evidence for the first time that merging clusters of 
galaxies are a promising candidate for the origin of high 
galactic-latitude, steady unidentified EGRET gamma-ray sources. 
Instead of using past optical catalogs of eye-selected clusters, 
we made a matched-filter survey of galaxy clusters over 
$4\arcdeg \times 4\arcdeg$ areas around seven steady unidentified 
EGRET sources at $|b|>45\arcdeg$ together with a 100 $\sq \arcdeg$ 
area near the South Galactic Pole as a control field. In total, 154 
Abell-like cluster candidates and 18 close pairs/groups of these 
clusters, expected to be possibly merging clusters, were identified 
within estimated redshift $z_{est}\leq 0.15$. Five among the seven 
EGRET sources have one or two cluster pairs/groups (CPGs) within 
1$\arcdeg$ from them. We assess the statistical significance of this 
result by several methods, and the confidence level of the real 
excess is maximally 99.8\% and 97.8\% in a conservative method. 
In contrast, we found no significant correlation with single clusters. 
In addition to the spatial correlation, we also found that the 
richness of CPGs associated with EGRET sources is considerably larger 
than those of CPGs in the control field. These results imply that a 
part of the steady unidentified EGRET sources at high-latitude are 
physically associated with close CPGs, not with single clusters. 
We also discuss possible interpretations of these results. We argue 
that, if these associations are real, they are difficult to explain 
by hadronic processes, but best explained by the inverse-Compton 
scattering by high energy electrons accelerated in shocks of cluster 
formation, as recently proposed. 

\end{abstract}


%----------------------------------------------------------------------

\section{Introduction}

The {\it Energetic Gamma Ray Experiment Telescope} (EGRET) aboard the 
{\it Compton Gamma Ray Observatory} ({\it CGRO}) has left us the third 
EGRET (3EG) catalog \citep{Har99}, the largest and deepest catalog of 
high-energy gamma-ray sources to date. 
However, more than 60\% of the 3EG gamma-ray sources (170 out of 271) 
are yet to be identified, mainly because of the poor accuracy of the 
position determination. 
The distribution of the unidentified EGRET sources can be explained as 
the sum of the Galactic ($|b| \la 40\arcdeg$) component and another 
isotropic (likely extragalactic) component \citep{Muk95, Oze96}. 
While several candidates were proposed for the origin of Galactic 
sources including molecular clouds, supernova remnants, massive stars, 
and pulsars \citep[e.g., see][and references therein]{Geh99}, no 
astronomical object except for AGNs has been proposed as the origin of 
the isotropic component consisting of about 20 sources at 
$|b| > 45\arcdeg$ ($\sim $ 65 in the whole sky). 
All AGNs identified as EGRET sources belong to the blazar class, and 
there is no evidence that other types of AGNs are emitting gamma-rays 
detectable by EGRET. 

Clusters of galaxies have been studied as a possible source of high 
energy gamma-rays, since high energy cosmic rays are expected to 
exist in intracluster medium (ICM), which could be emitted by member 
galaxies, or could be generated by AGNs or shocks in cosmological 
structure formation. Most previous studies concentrated on the 
hadronic processes, i.e., pion-decay gamma-rays produced by 
interaction between cosmic ray protons/hadrons with intracluster 
matter \citep{Vol96, Ber97, Col98}, and predictions are well below 
the detection sensitivity of the EGRET even for the case of the Coma 
cluster, for which only an upper limit has been set by the EGRET 
\citep{Sre96}. However, attention to high energy emission from 
nonthermal electrons is recently increasing. Existence of high energy 
electrons in intracluster medium has been suggested by diffuse 
nonthermal hard X-ray emission and diffuse radio emission for several 
clusters \citep[see e.g.,][for a review]{Sar01}. \citet{Loe00} pointed 
out that the extragalactic gamma-ray background in the GeV band may be 
explained by the inverse-Compton (IC) scattering of the cosmic 
microwave background (CMB) photons by electrons accelerated in 
large-scale shocks generated in structure formation, if about 5\% of 
the shock kinetic energy is converted into nonthermal electrons. 
\citet[hereafter TK]{Tot00} calculated the expected gamma-ray source 
counts by this process, and found that a few tens of sources are 
expected above the EGRET sensitivity from nearby dynamically forming 
clusters, and a part of unidentified EGRET sources may be accounted 
for \citep[see also][]{Wax00}. The preheating of the intergalactic 
medium, which is inferred from X-ray luminosity versus temperature 
relation of clusters and groups, may severely suppress the gamma-ray 
background flux, but still about 10 massive forming clusters could 
remain as gamma-ray sources detectable by the EGRET \citep{Tot01}. 

TK estimated that ``gamma-ray clusters'' detectable by EGRET should 
have typical redshift of $\la 0.1$ and mass of $\sim 10^{15} M_{\sun}$. 
However, no statistically significant correlation between the 
unidentified EGRET sources and known clusters has been found. There are 
several possible reasons for this result. First, only a small fraction 
of clusters should be emitting gamma-rays by the process considered by 
TK, since the cooling time of electrons emitting high-energy gamma-rays 
is very short ($\sim 10^6$yr) and hence only clusters which are 
dynamically forming with active shocks can emit GeV gamma-rays. The no 
detection from the Coma cluster is thus explained. We may not observe 
gamma-rays even from clusters with merging signatures in X-ray or radio 
bands, which remains on a much longer time scales than the gamma-ray 
emission. It is also difficult to select the candidates of gamma-ray 
clusters from all unidentified EGRET sources, since the Galactic 
gamma-ray sources extend to relatively high galactic latitude of 
$|b| \sim 45^\circ$ \citep{Geh00}, and a part of sources at even higher 
latitude seem to be variable and hence they are likely to be AGNs. Even 
if the candidates are appropriately selected, the typical redshifts 
reached in the existing all-sky cluster catalogs in optical 
\citep{Abe89} or X-ray \citep{Ebe98} are not much greater than the 
expected redshift of gamma-ray clusters ($z \sim 0.1$), and hence a 
part of gamma-ray clusters detected by EGRET could have been missed by 
the past cluster surveys. These facts make it difficult to search the 
correlation efficiently. Furthermore, TK pointed out that, since 
gamma-rays can be emitted only from just dynamically forming clusters, 
their structure may be considerably different and extended when 
compared with stable, well-established clusters detected by X-rays or 
optical surveys. This effect might make the correlation search with 
known clusters even more inefficient. 

However, most of such forming gamma-ray clusters should have some 
sub-structure or merging signature within them, as expected by the 
hierarchical structure formation in the CDM universe.\footnote{The 
terms ``forming'' and ``merging'' are difficult to clearly discriminate 
in the standard hierarchical structure formation, and hence we use them 
in essentially the same meaning.} Therefore, an intensive search for 
these signatures in the regions around the unidentified EGRET sources 
with sensitivities better than existing all-sky catalogs of galaxy 
clusters is a straightforward test of the gamma-ray cluster hypothesis.

In this paper, we report the first results from our project to 
systematically examine the gamma-ray cluster hypothesis using optical 
galaxy data.  Among the 19 unidentified EGRET sources at 
$|b| > 45\arcdeg$, we focus here on the 7 sources classified as 
``steady'' \citep[][and D.Macomb 2000, private communication]{Geh00} 
since the remaining 12 variable sources should be other objects such as 
flaring AGNs.  To perform a correlation analysis between the EGRET 
sources and galaxy clusters more efficiently than past studies, we make 
a new sample of galaxy clusters detected automatically based on the 
matched-filter cluster finding algorithm \citep{Kaw98}. This catalog 
should be better for statistical study of correlation, than the past 
optical cluster catalogs selected by eyes that inevitably induce some 
systematic bias. 
We found a statistically significant correlation at maximally 
$3.7 \sigma$ level between the 7 EGRET sources and close pairs of galaxy 
clusters, while no significant correlation was found with single clusters. 
We will argue that these results give an indirect support, though not 
conclusive, for the gamma-ray cluster hypothesis. 

%----------------------------------------------------------------------

\section{Data}

We use the galaxy sample extracted from the APM catalog. 
The data around the seven EGRET sources were obtained via APMCAT 
service\footnote{\url{http://www.ast.cam.ac.uk/\~{}apmcat/}} while the 
data near the SGP were kindly distributed from S. Maddox and M. Irwin. 
Only blue passband ($O$ or $b_J$) data have been used since the red 
passband data seemed much noisier for some EGRET source regions 
especially at the edge of photographic plates. 
Both $O$ and $b_J$ data were available and analyzed for 3EG J1235+0233. 
After correcting galaxy dimming due to Galactic absorption using 
the extinction maps and tools by \citet{Sch98}
\footnote{\url{http://astron.berkeley.edu/davis/dust/index.html}}, 
galaxies within magnitude range of $14 \leq m_{O,b_J} \leq 20$ were 
selected as the input data for cluster-finder.  The seven 
$4\arcdeg \times 4\arcdeg$ areas centered at each EGRET source are 
searched.  For comparison, a 100 $\sq \arcdeg$ area near the South 
Galactic Pole (hereafter SGP) is also used as a control field. Owing to 
the presence of holes and photographic plate edges in the data region, 
the total analyzed area is 182.93 $\sq \arcdeg$. 

%----------------------------------------------------------------------

\section{Cluster Identification}

To make an original cluster sample, we employed a revised version of 
the matched-filter method by \citet{Kaw98}, an automated and objective 
cluster-finding technique based on maximum-likelihood method. 
Here we briefly describe the essence of this revised matched-filter. 
%(full description will appear in a future paper). 
A likelihood value that a cluster is centered at a given point on the 
sky is computed using galaxies in a circular region with radius $r_{cir}$ 
centered at the point. The circular region is divided into five annular 
subregions and the galaxies in each subregion are then used to compare 
with the ``filter'', a model of spatial and magnitude distribution of 
cluster galaxies. Since we assume the King model for the surface density 
profile and the Schechter function for galaxy luminosity function, the 
``filter'' has several control parameters including cluster core radius 
$r_c$, shape parameter of King model $c$, Schechter parameters $M^*$ and 
$\alpha$, redshift $z_{fil}$, and richness ${\cal N}$ defined as the 
number of cluster galaxies brighter than $m^*+5$ and within Abell radius 
(=1.5$h^{-1}$ Mpc).  The relationship between ${\cal N}$ and Abell 
richness $C$ (the number of cluster galaxies within 2 magnitudes from the 
third brightest galaxy), obtained with Monte Carlo simulation, is given 
as $C = 1.1{\cal N}^{0.65}$ with uncertainty of 20\%. All parameters but 
$z_{fil}$, ${\cal N}$, and $r_c$ are fixed as $c = 2.25$, 
$\alpha = -1.25$, $M^*_{O} = -19.44 + 5\log h$, and 
$M^*_{b_J} = -19.8 + 5\log h$.  For {\sl K}-correction, the values for 
E/S0 galaxies by K.Shimasaku (2001, private communication) and 
\citet{Sha84} were used respectively for $O$ and $b_J$ passbands. 
Cosmological parameters are fixed as $h = H_0/100 = 0.8$ and $q_0 = 0.5$. 
The number and magnitude distribution of the foreground/background 
galaxies are locally estimated using the galaxies in an annular region 
around the point with the inner and outer radii of $0.5\arcdeg$ and 
$1\arcdeg$, respectively. 

In the first step of the actual procedure, we fix ($z_{fil}, r_c$) as 
(0.14, 50$h^{-1}$ kpc) and tune only ${\cal N}$ to maximize likelihood 
at a point in order to simplify calculation. 
Maximized likelihood and corresponding ${\cal N}$ are computed at lattice 
points with the interval of 0.01$\arcdeg$ to draw a ``likelihood map'' 
and a ``richness map''. 
We use the latter to detect clusters because of simpler appearance of 
clusters in ``richness map'' \citep[see Figure 2 of][]{Kaw98}. 
After smoothing the raw ``richness map'' with Gaussian filter with 
$\sigma = 0.03\arcdeg$, we detect cluster candidates appearing as local 
maxima with ${\cal N} > 161$ (i.e., Abell richness class $\geq 0$). 
Then $z_{fil}$ and $r_c$ are surveyed in the range of 
$0.04 \leq z_{fil} \leq 0.2$ and $10 \leq r_c \leq 400$ (in $h^{-1}$ 
kpc), respectively, to estimate redshift $z_{est}$ and richness 
${\cal N}_{est}$ for each candidate. 
To avoid erroneous estimation, the above procedure is run for four 
cases of different galaxy sampling with $r_{cir} = 0.05\arcdeg, 
0.1\arcdeg, 0.15\arcdeg, {\rm and} ~ 0.2\arcdeg$. 
Basically we adopt the values ($z_{est}, {\cal N}_{est}$) for the case 
$r_{cir} = 0.2\arcdeg$ unless they are far apart from the other values 
for the case $r_{cir} = 0.05\arcdeg, 0.1\arcdeg, {\rm and} ~ 0.15\arcdeg$. 
The uncertainty of $z_{est}$ is estimated to be $\sim 20\%$ with Monte 
Carlo simulation. 
Finally we obtain a volume-limited, ``three-dimensional'' sample of 
154 cluster candidates with Abell richness class $\geq 0$ complete 
out to $z=0.15$. 

Figure 1 shows central $2.8\arcdeg \times 2.8\arcdeg$ areas of the 
``richness maps'' around the seven EGRET sources. 
Cluster candidates are seen as local peaks of color contour. 
It should be noted that this color contour just indicates the amplitude 
of the ``best-fit'' filter with a {\it fixed} redshift parameter 
($z_{fil} = 0.14$) and does not directly reflect cluster's richness 
except for the ones at $z = 0.14$. 
Only the clusters with $z_{est} \leq 0.15$ and Abell richness class 
$\geq$ 0 (i.e., ${\cal N}_{est} > 161$), which we utilize below, 
are marked with the pluses. 

Using this cluster sample, we search for close cluster pairs or 
groups (hereafter CPGs) as candidates of merging clusters. 
If there are close clusters satisfying two criteria that (1) their 
estimated redshifts equal one another within the uncertainty of 
redshift estimation ($\sim 20\%$) and that (2) their transverse 
separation at that redshift is less than 2$h^{-1}$Mpc, we regard 
them as a CPG. 
In total, we identify 18 CPGs consisting of 2-4 clusters. 
Table 1 lists relative position (columns 3 and 4) and separation 
(column 5) from the nearest EGRET source, mean estimated redshift 
(column 6), total Abell richness (column 7), and number (column 8) 
of member clusters for the 9 CPGs found in the EGRET data areas. 
Some of the CPGs are shown as the green ellipses enclosing member 
clusters in Figure 1. 

%----------------------------------------------------------------------

\section{Results}

Here we try several statistical tests for the correlation between 
clusters and the 7 EGRET sources. 

\subsection{Projected Number Density}
\label{section:number}

We examine if there is an excess overdensity of clusters or CPGs in 
the vicinity of the EGRET sources (hereafter VES).  VES is defined as 
the sum of all area within 1$\arcdeg$ from the seven EGRET sources, 
and the boundary is shown as the yellow solid circles in Figure 1. 
Considering the extended nature of CPGs, the VES radius is fixed at 
1$\arcdeg$ rather than the EGRET error radius; the value is close to 
the typical size of both EGRET error circle and expected gamma-ray 
clusters detectable by EGRET (TK). The rest of the data area (the 
EGRET region outside VES plus the SGP region) is referred to as 
``control field'' hereafter. Considering the lack of galaxy data due 
to the photographic plate edges, VES and the control field cover 20.07 
$\sq \arcdeg$ and 162.86 $\sq \arcdeg$, respectively. 

Simply counting all clusters, VES and the control field contain 21 and 
133 clusters, respectively. The number of clusters expected by chance 
in VES should obey the Poisson distribution with the expectation value 
inferred from the control field, 133$\times $20.07/162.86 = 16.4 if we 
ignore cluster-cluster correlation. We see that there is only a weak 
density excess of clusters at 1.1 $\sigma$ level in VES. 

However, situation changes greatly for CPGs. 
Five among the seven EGRET sources, namely, all except for 3EG 
J1235+0233 and 3EG J1337+5029 have CPGs within 1$\arcdeg$ from them. 
Four EGRET sources (3EG J0038-0949, 3EG J1234-1318, 3EG J1310-0517, 
and 3EG J1347+2932) have one CPG and the other one (3EG J0159-3603) 
has two CPGs within 1$\arcdeg$. 
Anyway, there are 6 and 12 CPGs in VES and the control field, 
respectively. Therefore, the number of CPGs expected by chance 
in the 20.07 $\sq \arcdeg$ VES field is 
12$\times $20.07/162.86 = 1.5, thus the number excess of CPGs in VES 
amounts to $(6-1.5)/\sqrt{1.5} = 3.7 \sigma$ level (namely, 99.6\% CL 
assuming Poisson distribution), which is in sharp contrast to the case 
for single clusters. 
Even in a conservative case (increasing CPG number of the control field 
to $12+\sqrt{12}$, namely +1$\sigma$ level), the CPG number excess is 
at 3.0 $\sigma$ level (or 98.7\% CL). 
The weak correlation between single clusters and EGRET sources seems to 
appear under the influence of the strong correlation between CPGs 
and EGRET sources. 

\subsection{Mean Closest Separation}

Next we assess the correlation between CPGs and the EGRET sources 
in a slightly different way by computing the mean closest separation 
between CPGs and the EGRET sources and examining if it is smaller 
than that for the case if CPGs are randomly distributed. 
Using six EGRET sources except for 3EG J1337+5029, for which no CPG is 
found in the data area, the mean closest separation is 0.84$\arcdeg$. 
We then perform a Monte Carlo simulation to compute mean closest 
separation for random distribution case. 
We have 60000 realizations of random placement of CPGs with density 
of 12/162.86 = 0.074 /$\sq \arcdeg$ and then measure the distances 
of the closest CPGs from a given point. 
Computing mean of every six closest separations, we obtain the 
distribution of 10000 values of mean closest separation for random case. 
The mean and the standard deviation of this distribution are 
1.84$\arcdeg$ and 0.39$\arcdeg$, respectively. 
Using this distribution, the observed mean closest separation for the 
six EGRET sources is apparently smaller than that for random case with 
(1.84-0.84)/0.39 = 2.6 $\sigma$ level or 99.8 \% CL (for the 
conservative case in the previous subsection, 2.3 $\sigma$ level or 
99.5 \% CL). 
These results change only very little if we assume that there is a CPG 
just outside the $4\arcdeg \times 4\arcdeg$ area around 3EG J1337+5029.

\subsection{Bayesian Statistics Using Elliptical Fits}
\begin{figure}
\plotone{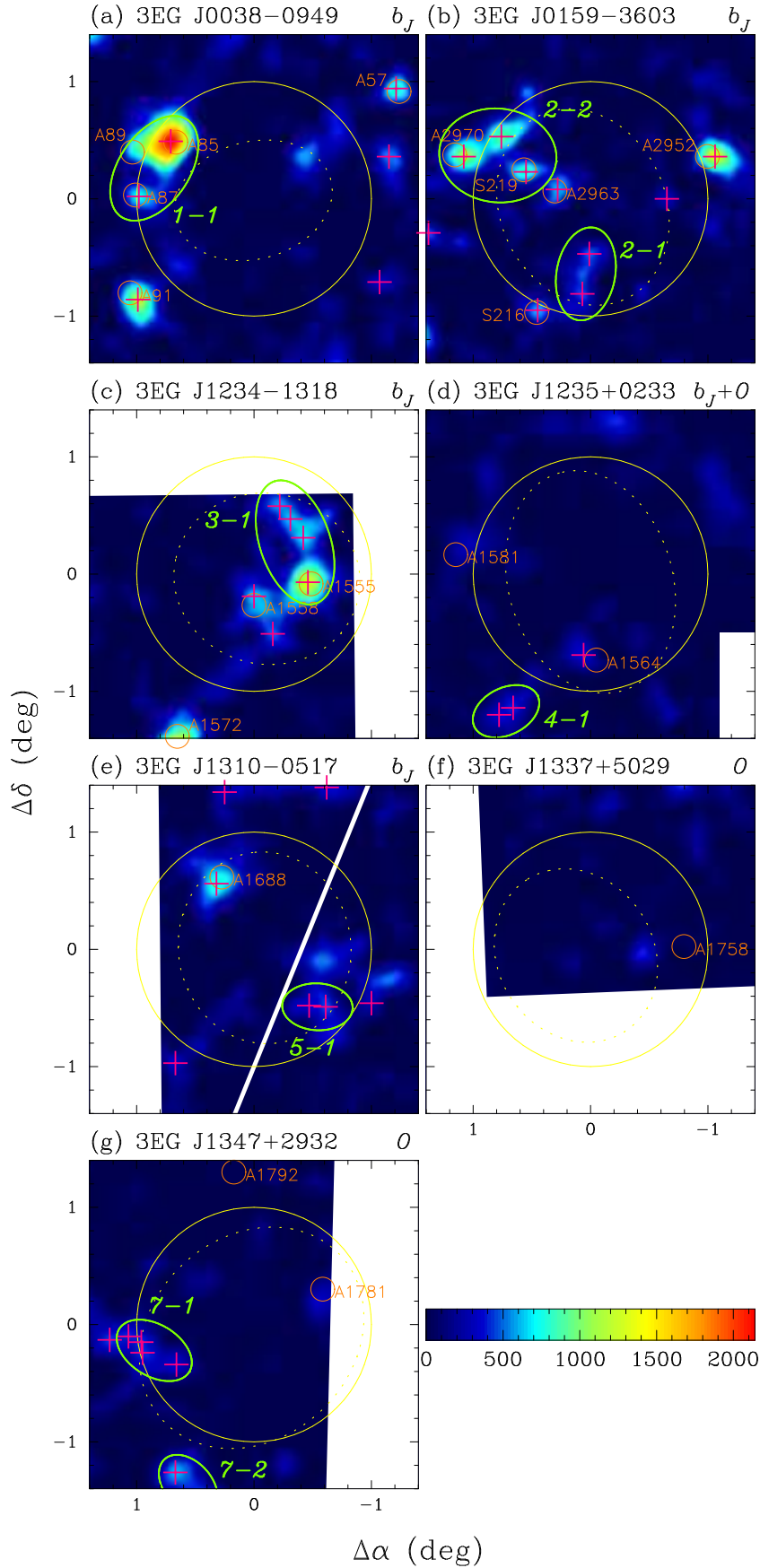}
\caption{
The Matched-Filter ``richness maps'' for the seven regions centered
at the steady unidentified EGRET sources at $|b|>45\arcdeg$. The EGRET source
name and the bandpass of the galaxy data are shown at the top of each
panel. The {\it pluses} denote the Abell-like cluster candidates with $z_{est}
\leq 0.15$ and Abell Richness Class $\geq 0$ detected by Matched-Filter. The
{\it small open circles} are Abell/ACO clusters for reference. Close cluster
pairs or groups (CPGs) are shown as the {\it green ellipses} enclosing their
member clusters. The boundary of VES is shown with the {\it large yellow
circles} (solid line). The {\it yellow dotted ellipses} denote the best-fit ellipses for 95\% confidence regions of the EGRET sources by \citet{Mat01}. 
}
\label{fig:map}
\end{figure}

\begin{figure}
\plotone{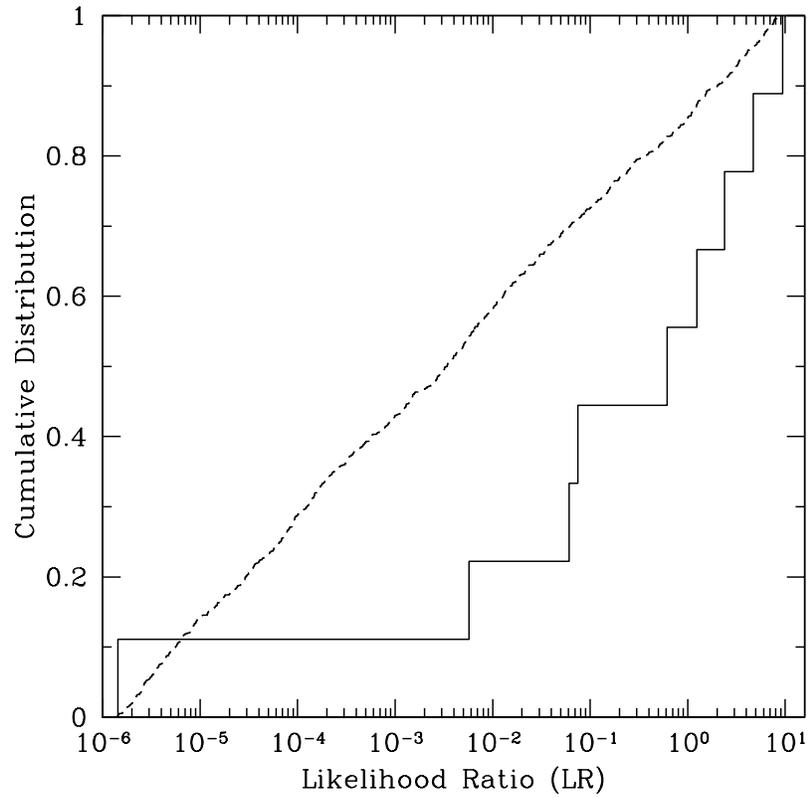}
\caption{Cumulative distribution of the likelihood ratio ($LR$)
of identification of CPGs as EGRET sources. The solid line is
for the 9 CPGs around the 6 EGRET sources considered in this paper,
which are listed in Table 1. The dashed line is the result of 
Monte Carlo simulation assuming no physical correlation between CPGs
and EGRET sources. The distribution is considered in a range of
$1.43 \times 10^{-6} < LR < 9.4$, which the range of $LR$ found for the
9 CPGs.
}
\label{fig:LR-dist}
\end{figure}

In addition to the rather simple-minded analyses in the previous two 
sections, we also performed more sophisticated correlation study based 
on the Bayesian statistics, with the same procedure that has been used 
in some past studies on EGRET source identifications 
\citep{Mat97, Mat01}. We can calculate the likelihood ratio of 
identification, $LR \equiv dp(r|id)/dp(r|c)$, for a CPG located at a 
separation angle of $r$ from the center of an EGRET source, where 
$dp(r|id)$ or $dp(r|c)$ are differential probabilities that a CPG is 
found at $r$ when the CPG is a correct identification of the EGRET 
source or a confusion noise, respectively. Here, the information of a 
mean CPG number density (12/162.86$\square^\circ$) and elliptical fits 
to the 95\% C.L. contour of the likelihood of the EGRET source location 
(shown by yellow dotted lines in Fig. \ref{fig:map}) are used to 
calculate $LR$ \citep[see][in detail]{Mat97}. The distribution of the 
likelihood ratio can be used as an empirical indication of the strength 
of a potential identification. The values of $LR$ are given for the 9 
CPGs in Table 1.\footnote{The relative positions of CPGs in Table 1 are 
from EGRET locations given in the 3EG catalog, while the centers of 
elliptical fits given by \citet{Mat01} are slightly different. We 
corrected this offset here.}  To compare with this distribution, we 
performed a Monte Carlo simulation (MC) to produce 900 random location 
of 9 CPGs (i.e., 100 per each) around the six EGRET sources, assuming 
no correlation between the two. 
Since $LR$ of the 9 CPGs is distributed in a range of $1.4 \times 
10^{-6}$--9.4, we compare the cumulative distribution of $LR$ to the MC 
in the same range, as shown in Figure \ref{fig:LR-dist}.  Clearly the 
distribution of the 9 CPGs is deviated towards higher $LR$ compared with 
that of the MC. The Kolmogorov-Smirnov test (KS) gives a chance 
probability of this deviation as 2.3\%, i.e., the observed $LR$ 
distribution is different from the MC with a confidence level of 97.7\%. 

Although this result also indicates the physical correlation between the 
CPGs and EGRET sources, the significance seems less than those estimated 
in the previous two sections. However, we should emphasize that the test 
in this section should be conservative because of the following reason. 
Due to the limited time to perform the matched-filter calculation, the 
search for CPGs is made only for regions surrounding EGRET error circles, 
and hence CPGs far from EGRET sources are excluded in the above sample 
of CPGs. Therefore we do not have any real CPG with 
$LR < 1.4 \times 10^{-6}$, and we have to compare the observed $LR$ 
distribution to the MC only in the limited range of 
$LR > 1.4 \times 10^{-6}$. This means that the absolute number of CPGs 
with $LR > 1.4 \times 10^{-6}$ is not taken into account in the 
statistical significance. On the other hand, the result of 
\S \ref{section:number} indicates that finding 9 CPGs with 
$LR > 1.4 \times 10^{-6}$ in the region around EGRET sources is higher 
than expected from random coincidence. Therefore the statistical 
significance only by the KS test in this section might be an 
underestimate. In addition, we took the separation $r$ to the center of 
CPGs, but it is uncertain where is the gamma-ray emitting region in the 
extended region of CPGs. Therefore the calculation of likelihood ratio 
might be too strict. 

We can infer the a priori probability, $p(id)$, that each of the 9 CPG 
is a correct identification of EGRET sources, and the a posteriori 
probability, $p(id|r)$, that a CPG located at $r$ is the correct 
identification, as follows. Again following \citet{Mat97}, we can 
calculate $p(id|r)$ for each observed CPG when unknown $p(id)$ is 
specified. Then, we solve a self-consistent $p(id)$ such that the 
integral of $p(id|r)$ divided by the number of CPGs considered (=9) 
yields the assumed value of $p(id)$. We found $p(id)=0.275$ here, and 
$p(id|r)$ assuming this value of $p(id)$ is also given for every CPG in 
Table 1. According to this $p(id|r)$ estimate, we can calculate a chance 
probability that all CPGs are misidentification, i.e., 
\begin{equation}
p_c = \prod_{i=1}^9 [1 - p_i(id|r_i)] \ ,
\end{equation}
where the subscript $i$ runs over the 9 CPGs. We found this probability 
to be 2.2\%; i.e., at least one GPG is the correct identification with 
97.8\% CL. 

\subsection{Estimated Redshift and Richness}
\label{section:mass-z}

The bottom panel of the Figure 1 of TK shows that the redshift and mass 
of gamma-ray clusters detectable by EGRET is $\sim$ 0.1 and $\sim 10^{15} 
M_{\sun}$, respectively.  Both of them are roughly consistent with the 
estimated values of the CPGs in the vicinity of EGRET sources (see columns 
6 and 7 of Table 1). Noting for richness, column 7 of Table 1 shows that 
the most CPGs have total Abell richness of $C_{\rm total} > $ 100 (Abell 
richness class 2-4).  This means that they are quite massive systems with 
mass of $\sim 10^{15} M_{\sun}$. 

We also found that the 6 CPGs within the VES of EGRET sources seem to 
have larger $C_{\rm total}$ compared with those not associated with 
EGRET sources. The 6 CPGs have $C_{\rm total}$ = 79, 109, 128, 165, 206, 
and 217, which should be compared with those of 12 CPGs in the control 
field: 62, 67, 90, 91, 92, 99, 102, 110, 111, 114, 119, and 154. If CPGs 
are not related to EGRET sources at all, the distribution of richness 
should be the same for the EGRET region and the control field. The KS 
test indicates that the chance probability of getting this result from 
the same distribution function is 8.0\%. This is not very compelling 
only by itself, but it should be noted that this test is completely 
independent of the spatial correlation discussed in the previous three 
sections. If this result is added to the spatial correlation, 
significance would be further increased. 

%----------------------------------------------------------------------

\section{Discussion}
\subsection{Variability of EGRET Sources}
In this work we selected 7 sources at $|b|>45^\circ$ that are showing 
no evidence of variability, according to \citet{Geh00}. However, the 
variability of EGRET sources cannot be determined clearly for many cases. 
In fact, there are two other studies on the variability by \citet{Tom99} 
and \citet{Tor01}, and the classification of EGRET sources into variable 
or non-variable sources is sometimes different among these authors. We 
also checked the variability indicators defined by Tompkins ($\tau$) and 
Torres et al. ($I$) for the seven EGRET sources here. According to the 
plausible criteria given by Torres et al. for these two indicators, they 
can be classified into either of ``variable'', ``dubious'', and 
``non-variable'' sources. We found that all but one of them are 
classified as non-variable or dubious sources in both the two indicators; 
however, only one source, 3EG J1310-0517 is classified as a variable 
source by the $\tau$ indicator, while it belongs to non-variable sources 
by the $I$ indicator. The difference seems to come from the analysis of 
Tompkins et al. utilizing EGRET data that are not included in the 3EG 
catalog, while the classification by Gehrels et al. or Torres et al. is 
based only on the 3EG catalog (P.L.Nolan 2001, a private communication). 

Considering this point, we also give statistical significance of 
correlation when 3EG J1310-0517 is removed from the sample. 
The number of CPGs expected by chance within the 6 VES of EGRET sources 
(18.15$\square^\circ$) is 1.33, and hence the observed 5 CPGs are 
$(5-1.33)/\sqrt{1.33} = 3.2\sigma$ excess of random coincidence 
(98.8\% CL in Poisson distribution). 
The KS chance probability of the likelihood ratio distribution in the 
Bayesian analysis becomes 5.1\%, and the chance probability that all CPGs 
are misidentification becomes $p_c = $ 7.1\%. 

\subsection{Theoretical Interpretation and Comments on the Other Work}
After the submission of this paper, we learned a recent study of 
\citet[hereafter C02]{Col02} who investigated the correlation between 
unidentified EGRET sources and Abell clusters. Our analysis is based on 
the newly produced cluster catalog based the automated matched-filter 
method, which is more reliable and objective for statistical cluster 
study than the eye-selected Abell clusters. On the other hand, C02 also 
examined radio and X-ray fluxes of clusters associated with unidentified 
EGRET sources. C02 found interesting correlations between X-ray, radio 
and EGRET gamma-ray fluxes that further strengthen the possible 
connection between clusters and EGRET sources. All but one 
(3EG J1235+0233) EGRET sources considered here are also included in the 
list of candidates selected by C02. Therefore, at first glance, 
observational results seem consistent with ours. However, it should be 
noted that only about half of the CPGs presumably associated to the 
EGRET sources have Abell clusters as their members. We also found no 
statistically significant excess of number density for single clusters. 
The correlation claimed by C02 is between single clusters and a larger 
number of EGRET sources at $|b|>20^\circ$ including variable sources 
than considered here. Such a correlation could also be induced by point 
sources (e.g., AGNs) residing in galaxy clusters. On the other hand, our 
result that only CPGs show strong correlation with steady EGRET sources 
indicates that origin of gamma-ray emission is shocks by 
cluster/structure formation. 

The theoretical interpretation of these results by C02 is very different 
from ours; in fact, he strongly argued that the forming/merging clusters 
proposed by TK are not responsible for the association suggested by this 
work and/or C02. Here we give a detailed interpretation of our results 
giving some comments on C02's arguments against forming/merging clusters, 
and argue that the suggestion made by TK is the best explanation of the 
possible association between CPGs and EGRET sources. 

To begin with, let us make clear what are the essentially new aspects of 
the proposal by TK: this work considered the IC scattering by electrons 
accelerated in shocks generated by the process of hierarchical cluster 
formation. This work is the first to predict gamma-ray source counts 
expected by such process based on the standard structure formation 
theory in the CDM universe, and TK found that maximally a few tens of 
forming clusters could be detectable by the EGRET. On the other hand, 
previous studies concerning gamma-rays from galaxy clusters mostly 
considered the hadronic processes such as pion decays by primary cosmic 
ray protons and emission from secondary electrons 
\citep{Vol96, Ber97, Col98}. 
Generally these papers found gamma-ray fluxes well under the EGRET 
sensitivity limit, even for a sample of the closest clusters to us 
including Coma \citep{Col98}. The energy loss time scale of high energy 
protons in clusters is comparable to, or longer than the Hubble time, 
and hence the gamma-ray luminosity should also be steady on this time 
scale. Then, there is no reason to expect even stronger gamma-ray flux 
from clusters other than Coma. Therefore gamma-rays produced by cosmic 
ray protons in galaxy clusters have not been seriously considered as a 
candidate of unidentified EGRET sources. 

However, high energy electrons that can emit gamma-rays have much 
shorter cooling time scale ($t_{\rm cool} \sim 10^6$yr) than that of 
protons or ions. Therefore, if a comparable energy is going into 
cosmic-ray electrons and protons, then we expect much stronger gamma-ray 
luminosity of the electron origin when the shock is still active after 
formation or merging processes, since their energy is emitted within a 
duration of shock life time, i.e., dynamical time ($\sim$ Gyr). (Note 
that we should not use $t_{\rm cool}$ here.) Furthermore, we expect 
gamma-rays only from clusters still having active shocks, and do not 
from well stabilized clusters without shocks. The gamma-ray flux is 
expected to vary strongly in the history of hierarchical formation of a 
cluster.  Therefore, it is possible that Coma and other nearby clusters 
do not have strong gamma-ray emission, while other more distant, or less 
stabilized clusters emit gamma-rays detectable by EGRET. 
What TK found is that it is in fact quantitatively possible, based on 
the abundance of forming objects in the CDM universe. 

C02 first criticized an inconsistency that TK claimed forming clusters 
with undetectable X-ray flux or galaxy clustering due to more extended 
profile than normal clusters, while this work claims the correlation 
between EGRET sources and cluster pairs as the support of the TK's idea. 
It should be noted, however, that only 3 of the 8 CPGs shown in 
Fig. \ref{fig:map} are coincident with Abell clusters. This suggests that 
a significant part of CPGs found by our paper are very extended and only 
found as clustering of galaxy clusters, each of which is small and not in 
the Abell catalog. (However, as shown in \S \ref{section:mass-z}, the 
total mass of these CPGs is as large as $10^{15}M_\odot$.)  This is in 
fact consistent with the picture of TK.  On the other hand, coincidence 
with known Abell clusters for some of CPGs and EGRET sources is also not 
surprising, since, in the hierarchical strucuture formation theory, 
forming or merging clusters sometimes should include rich clusters that 
can be detected by past surveys.  One important point is that, even in 
such cases, TK predicts that gamma-ray emission should not be from the 
center of rich clusters. Rather, gamma-ray emission is expected from more 
extended region of CPGs including the rich clusters. 

We note that \citet{Rei99} set an upper limit on the gamma-ray flux from 
A85 as $<6.77 \times 10^{-8} \ \rm cm^{-2} s^{-1}$ ($>$100MeV), instead 
of accounting for the nearby unidentified source, 3EG J0038-0949 (On the 
other hand, this association is classified as the most probable 
association in C02). If gamma-ray flux is coming from hadronic processes,
then we expect that the flux should be the strongest at the center of 
A85, where cosmic-rays are well confined and ICM density is the highest. 
Therefore the hadronic processes cannot explain the association of A85 
and 3EG J0038-0949. On the other hand, as discussed by TK, the IC 
gamma-ray emission from forming or merging clusters should be more 
extended because of more extended ICM and uniform density of the CMB. 
Formation shocks are expected in the surrounding region of the CPG 
including A85, which seems marginally overlapping with the 95\% ellipse 
of 3EG J0038-0949. Therefore, if A85 and 3EG J0038-0949 are physically 
associated, the IC gamma-rays should be a better explanation than 
hadronic processes. 

Second, C02 claimed apparent discrepancy between the numbers of clusters 
in TK and this work: 20--50 predicted by TK and 7 found in this work. 
Here C02 did not take into account the sky coverage; 20--50 of TK is for 
all sky but 7 in this work is for $|b|>45^\circ$. The correction factor 
makes these numbers consistent. We also note that 20--50 sources 
predicted by TK might be a rather optimistic value, since electron power 
index is assumed to be $\alpha = 2$ ($dN_e/dE_e \propto E_e^{-\alpha}$). 
Somewhat softer spectrum is expected in reality \citep[see][]{Tot01}, 
reducing detectable sources by EGRET. However, it should be noted again 
that even such reduced number of detectable sources is still much larger 
than that expected by the hadronic processes considered before TK. 

Third, C02 claimed the difficulties of energetics in the theory of TK; 
the gamma-ray luminosity inferred from EGRET sources is much larger than 
that possible in the TK's framework. We again emphasize that TK assumes 
that 5\% energy injection from shock kinetic energy to nonthermal 
electrons. The kinetic energy is calculated based on the standard 
structure formation theory. It is a general thought that supernova 
remnants inject about 10\% of explosion energy into cosmic ray protons, 
although it may seem relatively large for electrons when one considers 
the energetics ratio of $\sim 0.01$--0.1 of cosmic ray electrons to 
protons observed at the Earth. However, radio observations of supernova 
remnants indicate that shocks convert at least a few percent of the 
shock energy into the acceleration of relativistic electrons 
\citep{Bla87, Sar01}. The EUV and hard X-ray emission from several 
clusters can be attributed to nonthermal electrons having the total 
energy of the same order \citep{Fus99, Sar01}. Therefore, the energetics 
assumed by TK is not extreme at all. C02 ignored this fundamental point, 
and showed discrepancies in some quantities assuming cluster mass of 
$10^{14} M_{\odot}$. It is not clear why C02 chose this very small mass 
instead of the standard value of $10^{15} M_\odot$. 
Since $L_\gamma \propto M^2$ in eq. (7) of C02, this choice reduces 
the gamma-ray flux by a factor of 100. On the other hand, TK's 
calculation shows that the typical mass of clusters that are detectable 
by EGRET is in fact $\sim 10^{15}M_\odot$. The richness estimate of 
CPGs in this paper also indicates similar masses 
(see \S \ref{section:mass-z}). Therefore, it is not surprising that C02 
found some discrepancy, but it does not give any argument against TK. 
C02 also claimed that TK's scenario results in extraordinary temperature 
of intracluster matter, $\sim$ 27--270keV, but it again seems to 
originate from nonstandard choice of C02 for cluster parameters 
[$T=8$keV for $M=10^{14}M_\odot$ that is not supported by the $M$-$T$ 
relation of observed clusters \citep[e.g.,][]{Fin01}, and small density 
of $n = 10^{-4} \rm cm^{-3}$ rather than a typically used value of 
$10^{-3} \rm cm^{-3}$]. TK predicts that temperature of clusters 
detectable by EGRET should not be much different from that of normal 
clusters. Rather, the temperature could be sometimes lower if the shock 
kinetic energy has not yet been dissipated well in most of the 
intracluster medium. 

Forth, C02 claimed that there is no evidence for strong ongoing shocks 
in the sample presented in this work, while the model of TK predicts 
strong nonthermal emission in hard X-ray and extreme ultraviolet (EUV) 
bands by the IC scattering, and in the radio band by the synchrotron 
emission, by the same electron population producing gamma-rays. TK model 
indeed predicts hard X-ray and EUV emission at a similar flux (in 
$\nu F_\nu$) to that in gamma-ray band in most cases\footnote{This may 
not be the case in the very early stage of merging; see the next 
paragraph.}. However, as repeatedly mentioned by C02, there is almost no 
observational information in these bands for the sample of EGRET sources 
considered in this paper. No evidence by no observation is trivial, and 
it argues against neither of TK nor this paper. Instead, intensive 
follow-up observations for the sample presented here in hard X-ray and 
EUV bands would give an important test of the TK's scenario. 
The synchrotron radio flux depends sensitively on the strength of 
magnetic field; the $\nu F_\nu$ flux should scale as 
$\nu F_\nu({\rm radio})/\nu F_\nu({\rm gamma}) \sim 
U_{\rm mag}/U_{\rm CMB}$, where $U_{\rm mag}$ and $U_{\rm CMB}$ are the 
energy density of magnetic field and CMB, respectively. When the cluster 
magnetic field is at a level of $\sim 3\mu$G ($U_{\rm mag} \sim 
U_{\rm CMB}$), we expect similar strong flux in radio bands. However, 
observed hard X-ray flux and radio flux from the Coma cluster indicates 
$\sim 0.15\mu$G for this cluster \citep{Fus99, Sar01}, and then the radio 
flux should be 0.25\% of gamma-ray flux. Considering also that there is 
no deep radio observations for the sample in this paper, no evidence for 
strong radio emission does not necessarily contradict with TK's suggestion.

C02 noted that A85 in the vicinity of 3EG J0038-0949 is associated with 
cold front that is a possible signature of early stage of merging. C02 
claimed that this is not an ongoing violent merging processes, and hence 
this argues against TK. However this is not the case. A85 has a radio halo 
found on the border between substructures, where the cluster gas is first 
being shocked \citep{Sar01}. Once the merging starts and shocks are 
generated, the IC gamma-ray flux should rapidly increase with a minimum 
time scale of cooling and acceleration of high energy electrons ($\sim 
10^6$yr, TK). After this situation is achieved, gamma-ray flux is expected 
to be rather steady during the shock is propagating. On the other hand, 
nonthermal radiation in other wavebands (X, EUV, and radio) will achieve 
this steady state at a time scale of energy dissipation of responsible 
particles. Since this time scale is longer than that for gamma-ray band, 
the initial rise of luminosity could be slower than in gamma-rays. 
Therefore, strong gamma-ray flux is theoretically expected even when 
observations in other wavebands show evidences only for early merging 
stage. 

C02 concluded that ``the energy release at gamma-ray energies $E > 100$MeV 
of the EGRET-cluster associations is probably due to a superposition of 
diffuse (associated with the active ICM of the cluster) and concentrated 
(associated with the active galaxies living within the cluster) gamma-ray 
emission''. We do not disagree with this statement; our claim is that the 
physical process responsible for ``the active ICM of the cluster'' should 
be IC scattering by primary cosmic-ray electrons produced by structure 
formation, which is the central point of TK's proposal.  On the other 
hand, other processes within the standard physics, such as hadronic 
processes, are unlikely to explain gamma-ray flux detectable by EGRET 
from active ICM, as mentioned above. In fact, based on the hadronic 
processes, \citet{Col98} predicted gamma-ray flux much smaller than the 
EGRET sensitivity limits for nearby clusters at $z =$ 0.01--0.07. The 
majority of Abell clusters that are claimed to be associated with EGRET 
sources by C02 have even larger distances of $z \gtrsim $ 0.1 (see Table 
1 of C02). Since the gamma-ray flux by hadronic processes is not expected 
to vary significantly from cluster to cluster because of the long 
dissipation time scale, it seems difficult to explain the gamma-ray flux 
from ICM by the Colafrancesco \& Blasi model. 

\section{Conclusions}

We performed a correlation analysis between the 7 steady unidentified 
EGRET sources in the high-latitude sky ($|b|>45\arcdeg$) and a 
quasi-three-dimensional catalog of galaxy clusters newly generated with 
a matched-filter algorithm.  While there is no correlation between the 
EGRET sources and the individual clusters, in sharp contrast, we found 
a strong (maximally 99.8\%CL level) correlation between the EGRET sources 
and close pairs/groups (CPGs). This result is consistent with the 
gamma-ray cluster hypothesis proposed by \citet{Tot00}, which expect 
that the gamma-ray emission comes only from ongoing mergers with active 
shocks, but not from usual ones in dynamically ``quiet'' regime where 
the violent shock has subsided.  Because of the short time scale of 
energy dissipation, gamma-ray luminosity should more rapidly rise with 
the generation of the shock, and decay with the disappearance of the 
shock, compared with the thermal or nonthermal emission in longer 
wavelength.  This suggests that some clusters may have strong gamma-ray 
emission with still weak or only early signatures of merging in other 
bands, and others may have weak gamma-ray flux but with still remaining 
merging signature in longer wavelength. Confirmation of merging 
signatures in CPGs found in this paper is important for further 
verification, but deep observation is necessary when the merging is 
still in the early stage. 
%Detailed mapping of gravitational potential in these ``CPG'' 
%region by redshift survey of galaxies or weak gravitational 
%lensing analysis might also help to confirm if the ``CPG'' is 
%really a merging system. 

Clearly, the weak point of our analysis is the small sample (seven) of 
the steady unidentified gamma-ray sources due to the flux limit of 3EG 
catalog (though it is deepest to date). 
However, TK predicted that future gamma-ray telescope such as GLAST 
could find hundreds to thousands of gamma-ray clusters up to $z=$0.2-0.3. 
The coming three-dimensional deep galaxy catalogs from ongoing SDSS 
and 2dF survey projects will be ideal resources to directly compare 
with the GLAST gamma-ray sources. 
When it is established that a part of the extragalactic steady gamma-ray 
sources are from forming (merging) clusters, large-scale distribution 
of gamma-ray clusters will offer unique and valuable information about 
the dynamical side of cosmological structure formation, in contrast to 
the more stationary side that has been probed by conventional galaxy 
clusters in X-ray and optical bands. 

%----------------------------------------------------------------------

\acknowledgments

We are grateful to K.Shimasaku for computing {\sl K}-correction for $O$ 
passband and $O-b_J$ color of present E/S0 galaxies. W.K. is supported 
in part by Japan Society for the Promotion of Science (JSPS) Research 
Fellowship. T.T. is supported in part by the JSPS Postdoctoral Fellowship 
for Research Abroad (2001). 

%----------------------------------------------------------------------
%----------------------------------------------------------------------

%----------------------------------------------------------------------
%----------------------------------------------------------------------
\begin{deluxetable}{clrrccrcccc}
\tabletypesize{\scriptsize}
%\small
\tablewidth{0pt}
\tablenum{1}
\tablecolumns{8}
\tablecaption{Cluster Pairs/Groups (GPGs) in the EGRET data areas}
\tablehead{
ID & EGRET Source & $\Delta \alpha (\arcdeg )$ & $\Delta \delta (\arcdeg )$ &
Separation (\arcdeg ) & $\overline{z}$ & $C_{total}$ & $N_{cl}$ 
& $LR$ & $p(id|r)$ & Note }
\startdata 
1-1 & 3EG J0038$-$0949 &  0.85 &  0.26 & 0.88 & 0.055 & 128 & 2 & 6.6E-2 &
  2.2E-2 & \\
2-1 & 3EG J0159$-$3603 &  0.04 & -0.64 & 0.64 & 0.141 & 109 & 2 & 4.7 & 0.64
  & \\
2-2 & 3EG J0159$-$3603 &  0.79 &  0.37 & 0.88 & 0.116 & 206 & 3 & 6.1E-1 &
  0.19 & \\
2-3 & 3EG J0159$-$3603 & -1.71 & -0.20 & 1.72 & 0.104 & 111 & 2 & 1.43E-6 &
  5.5E-7 & outside VES \\
3-1 & 3EG J1234$-$1318 & -0.35 &  0.32 & 0.48 & 0.101 & 217 & 4 & 9.4 & 0.78
  & \\
4-1 & 3EG J1235$+$0233 &  0.72 & -1.17 & 1.37 & 0.071 &  67 & 2 & 5.7E-3 &
  2.2E-3 & outside VES \\
5-1 & 3EG J1310$-$0517 & -0.54 & -0.49 & 0.73 & 0.092 &  79 & 2 & 2.4 & 0.48
  & \\
7-1 & 3EG J1347$+$2932 &  0.91 & -0.21 & 0.93 & 0.044 & 165 & 4 & 1.2 & 0.32
  & \\
7-2 & 3EG J1347$+$2932 &  0.56 & -1.37 & 1.48 & 0.054 &  62 & 2 & 7.5E-2 &
  2.7E-2 & outside VES \\
\enddata

\tablecomments{Col.(1): IDs corresponding to those in Fig. \ref{fig:map}.
Col.(3)--(5): Relative celestial coordinate and separation from the central
position of the 3EG catalog. Col.(6): Mean of estimated redshift for CPGs.
Col.(7): Total Abell richness. Col.(8): Number of cluster members included in
a CPG.  Col.(9): Likelihood ratio of identification as an EGRET source.
Col. (10): The a posteriori probability of identification. }
\end{deluxetable}
%----------------------------------------------------------------------

%----------------------------------------------------------------------

\epsscale{0.6}

%----------------------------------------------------------------------
%----------------------------------------------------------------------

%----------------------------------------------------------------------

\end{document}